\documentclass[aps,prappl,reprint,a4paper,superscriptaddress,floatfix,amsmath,amssymb,amsfonts,noshowpacs,longbibliography]{revtex4-2}
\usepackage{babel}
\usepackage[bookmarksopen,bookmarksnumbered]{hyperref} 
\usepackage{natbib}     

\usepackage{newtxtext,newtxmath}
\usepackage[T1]{fontenc}
\usepackage[utf8]{inputenx}
\input{ix-utf8enc.dfu}
\usepackage{graphicx}
\usepackage[dvipsnames]{xcolor}
\usepackage{soul}
\usepackage[textwidth=17.5cm,textheight=23.5cm,verbose,pdftex]{geometry}

\begin{document}

\title{Enabling two-dimensional electron gas with high room-temperature electron mobility exceeding 100~cm$^2$/Vs at a perovskite oxide interface}


\author{Georg Hoffmann}
\email[Electronic mail: ]{hoffmann@pdi-berlin.de}
\affiliation{Paul-Drude-Institut für Festkörperelektronik, Leibniz-Institut im Forschungsverbund Berlin e.\,V., Hausvogteiplatz 5--7, 10117 Berlin, Germany}
\author{Martina Zupancic}\affiliation{Leibniz-Institut für Kristallzüchtung,Max-Born-Straße 2, 12489 Berlin}
\author{Aysha A. Riaz}\affiliation{Department of Chemistry, Imperial College London, London, UK}
\author{Curran Kalha}\affiliation{Department of Chemistry, Imperial College London, London, UK}
\author{Christoph Schlueter}\affiliation{Deutsches Elektronen-Synchrotron DESY, Hamburg, Germany.
}
\author{Andrei Gloskovskii}\affiliation{Deutsches Elektronen-Synchrotron DESY, Hamburg, Germany.
}
\author{Anna Regoutz}\affiliation{Department of Chemistry, Imperial College London, London, UK}
\author{Martin Albrecht}
\affiliation{Leibniz-Institut für Kristallzüchtung,Max-Born-Straße 2, 12489 Berlin}
\author{Johanna Nordlander}
\affiliation{Paul-Drude-Institut für Festkörperelektronik, Leibniz-Institut im Forschungsverbund Berlin e.\,V., Hausvogteiplatz 5--7, 10117 Berlin, Germany}
\author{Oliver Bierwagen}
\affiliation{Paul-Drude-Institut für Festkörperelektronik, Leibniz-Institut im Forschungsverbund Berlin e.\,V., Hausvogteiplatz 5--7, 10117 Berlin, Germany}


\begin{abstract}
In perovskite oxide heterostructures, bulk functional properties coexist with emergent physical phenomena at epitaxial interfaces. Notably, charge transfer at the interface between two insulating oxide layers can lead to the formation of a two-dimensional electron gas (2DEG) with possible applications in, e.g.,\ high-electron-mobility transistors and ferroelectric field-effect transistors. So far, the realization of oxide 2DEGs is, however, largely limited to the interface between the single-crystal substrate and epitaxial film, preventing their deliberate placement inside a larger device architecture. Additionally, the substrate-limited quality of perovskite oxide interfaces hampers room-temperature 2DEG performance due to notoriously low electron mobility. In this work, we demonstrate the controlled creation of an interfacial 2DEG at the epitaxial interface between perovskite oxides BaSnO$_3$ and LaInO$_3$ with enhanced room-temperature electron mobilities up to 119 cm$^2$/Vs -- the highest room-temperature value reported so far for a perovskite oxide 2DEG. Using a combination of state-of-the-art deposition modes during oxide molecular beam epitaxy, our approach opens up another degree of freedom in optimization and $in$-$situ$ control of the interface between two epitaxial oxide layers away from the substrate interface. We thus expect our approach to apply to the general class of perovskite oxide 2DEG systems and to enable their improved compatibility with novel device concepts and integration across materials platforms.
\end{abstract}


\maketitle

\section{Introduction}
\label{sec:introduction}

The family of complex oxides comprises dielectric, semiconducting, superconducting, ferromagnetic, and ferroelectric materials. The wide range of functionalities hosted by especially the perovskite $AB$O$_3$ oxides, in combination with the possibility of integration on silicon \cite{wang2019}, motivates the realization of oxide electronic heterostructures with precision on par with established semiconductor thin-film synthesis. By interfacing such materials epitaxially, new (multi)functional heterostructures can be constructed exhibiting physical phenomena beyond bulk properties \cite{Hwang2012,Ramesh2019}. Notably, oxide interfaces, representing two-dimensional objects within a multilayer structure, can give rise to two-dimensional electron gases (2DEGs) \cite{ohtomo2004}, superconductivity \cite{reyren2007}, interface magnetism \cite{Brinkman2007}, and Rashba spin-orbit coupling \cite{Caviglia2010}. In the case of 2DEGs, non-polar SrTiO$_3$ (STO) interfaced with other formal-polar perovskites (exhibiting charged lattice planes), such as LaAlO$_3$ (LAO) \cite{janotti2012,mannhart2008,ohtomo2004}, LaTiO$_3$\cite{ohtsuka2010}, or GdTiO$_3$\cite{moetakef2011,mikheev2015}, served as a work-horse for polarization-discontinuity-doped 2DEGs that pushed the evolution of exotic physical phenomena such as spin-charge conversion \cite{noel2020} and 2D superconductivity \cite{reyren2007}. These systems suffer, however, from a low STO-related room-temperature (RT) electron mobility ($\mu_\text{RT}<10~$cm$^2/$Vs) \cite{mannhart2008, khan2017} limiting their applicability for most device types.

Other oxide systems have been proposed to realise higher room-temperature 2DEG mobility. One such system is based on BaSnO$_3$ (BSO), a wide-bandgap semiconducting oxide with the highest reported bulk (three-dimensional) room-temperature electron mobility of up to 320~cm$^2$/Vs in single crystals \cite{kim2012a} and 120~-~180~cm$^2$/Vs in La-doped BSO (LBSO) thin films grown by molecular beam epitaxy (MBE) \cite{paik_2017a,raghavan2016,prakash_2017}. Theoretical calculations suggest that interfacing BSO with lattice-matched LaInO$_3$ (LIO) lead to the formation of a 2DEG inside the BSO for an SnO$_2$/LaO terminated interface layer due to polar-discontinuity doping \cite{krishnaswamy2016}, while for the BaO/InO$_2$ interface termination a two-dimensional hole gas inside the LIO is formed \cite{aggoune2021}. 
Electrical transport measurements on these and other systems using BSO as a channel material validate the presence of a 2DEG, however, they are struggling with freeze-out of the 2DEG \cite{eom2022} or need additional La doping of the BSO channel \cite{kim_2016,kim2019,pfutzenreuter2022}. These studies result in RT mobility values of 40~-~60~cm$^2$/Vs and leave a key subject of interest untouched, i.e., interface termination.
A recent combined experimental and theoretical study indicates a preference for SnO$_2$/LaO terminated interface formation in pulsed laser deposition (PLD) and MBE grown BSO/LIO heterostructures - yet without electrical transport data \cite{zupancic2024}.
In comparison, References.~\cite{kim2022,kim2023a} show that the intentional deposition of an SnO$_2$ layer at the BSO/LIO interface results in a decrease of sheet resistance. However, charge carrier density (CCD) and mobility remain behind the values of the abovementioned systems by one order of magnitude.

\begin{figure*}[t]
\centering
\includegraphics[width=1.0\textwidth]{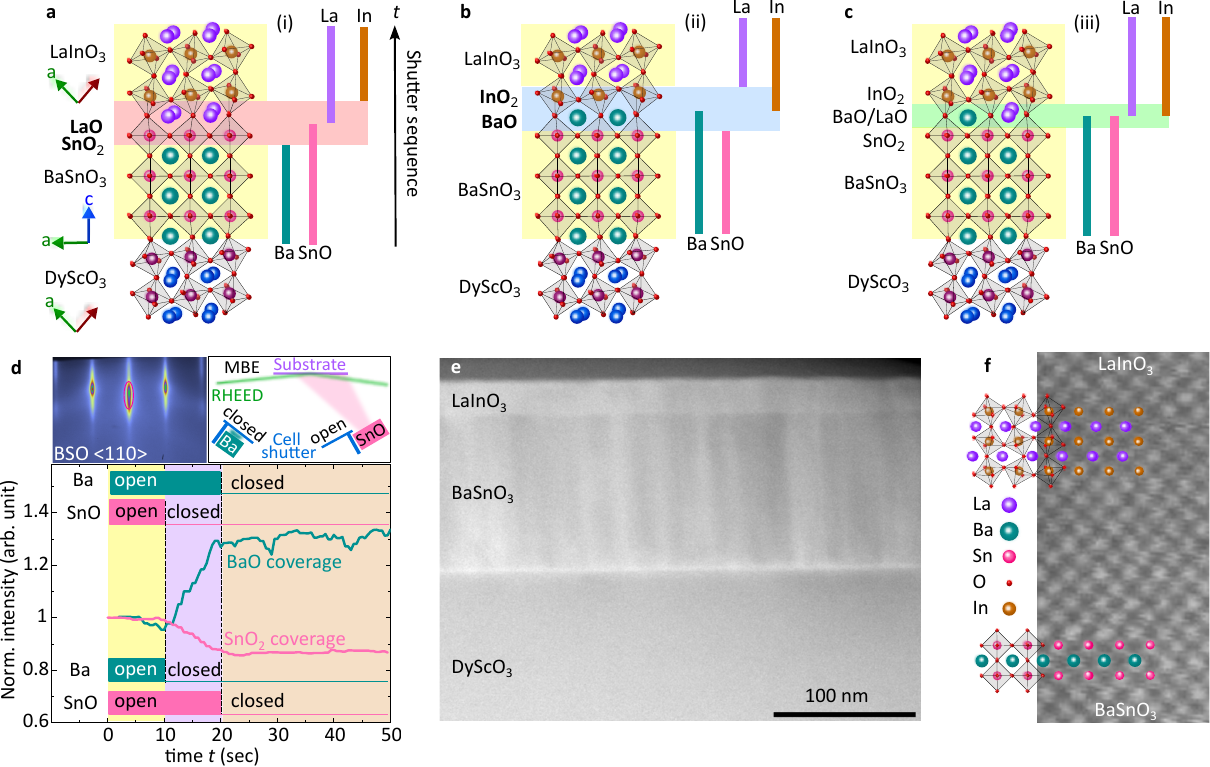}
\caption{\textbf{a--c} Sketch of the three different growth approaches to define the interface termination of a BSO/LIO heterostructure. The yellow shaded areas indicate co-deposition growth in an adsorption-controlled growth regime while the red (blue) shaded areas highlight interface termination using layer-by-layer growth. The vertical bars indicate the related shutter-controlled supply of particle flux from the corresponding effusion cell. \textbf{d} Normalized RHEED intensity of the specular spot (red tracer circle shown in the upper left corner) during BSO growth as a function of time followed by shutter-controlled deposition (see sketch in the upper right corner) for the realization of  SnO$_2$ (magenta curve) vs. BaO (blue curve) termination. 
\textbf{e} Cross-sectional STEM bright field image of the LIO/BSO/DSO heterostructure. \textbf{f} Enlarged area of the BSO/LIO interface revealing coherent growth of the heterostructure. Atomistic models of BSO and LIO were superimposed to the TEM image using VESTA \cite{momma2011}.}\label{fig_01}
\end{figure*}
STO-based 2DEGs\cite{mannhart2008} have been demonstrated on STO substrates whose surface termination has been accurately controlled by chemical preparation\cite{koster1998,ohnishi2004}. Yet, the issue of controlled interface termination away from the substrate remains an outstanding challenge for 2DEGs formed at oxide interfaces \cite{nakagawa2006,thapa2021}. The realization of a reproducible growth protocol for perovskite oxides, ensuring both the highest crystalline quality (maximizing electron mobility) and the most chemically sharp interfaces (maximizing charge transfer, hence carrier concentration), would open up for the insertion of 2DEG interfaces into intended positions in the device environment, since the substrate interface no longer participates.

While adsorption-controlled MBE, by co-deposition of cations A and B and self-regulated stoichiometry due to desorption of the excess cation, has demonstrated the highest material quality of ABO$_3$ perovskite thin films \cite{son2010,paik_2017a,raghavan2016,prakash2015}, it lacks a good definition of the surface termination (AO or BO$_2$) required for the heterostructure. In turn, the termination is best defined in a shutter-controlled layer-by-layer growth \cite{jalan2009,nie2014,wrobel2017}, in which an AO and a BO$_2$ monolayer (ML) are alternatingly deposited, yet coming at the expense of inevitably accumulating a net non-stoichiometry due to the limited control on the provided A and B fluxes. 

In the following, we identify control of the interface termination as a key ingredient for high-mobility 2DEGs in BSO/LIO heterostructures. For this purpose, we demonstrate a growth protocol of the interface termination by sequentially changing from adsorption-controlled MBE of BSO and LIO layers to layer-by-layer growth of their interfacial monolayers.
A full set of electrical transport data verifies that our approach of interface termination results in BSO-based 2DEGs with record RT mobilities up to 119~cm$^2$/Vs -- even in the absence of lattice matching to the substrate, indicating the potential for even further improvement of this value. The employed approach, which can be generalized to almost all perovskite oxide interfaces, has the potential to empower the position of BSO/LIO and similar perovskite heterostructure systems within the field of 2DEG-based high electron mobility transistors (HEMTs) and related devices \cite{jany2014,park2020}. 

\section{Results and Discussion}\label{sec:results}
\subsection{Designing the interface termination}\label{subsec_design}

To demonstrate the intentional control of the 2DEG formation by controlling the interface termination in BSO/LIO heterostructures, three distinctly different approaches to interface formation are compared during the transition from the growth of the BSO layer to the LIO layer on top. 
As schematically shown in Fig.~\ref{fig_01}\textbf{a--c}, these approaches are (i) the growth of a shutter-controlled SnO$_2$ monolayer (ML) followed by one ML of LaO, (ii) the growth of a shutter-controlled BaO ML followed by one ML of InO$_2$, or (iii) the absence of shutter controlled ML
deposition. In all three approaches, the bulk of the BSO and LIO layers are individually grown by adsorption-controlled co-deposition to benefit from the associated low rate of point defect formation in this growth mode \cite{paik_2017a,raghavan2016,prakash2015,hoffmann2022}.

 First, we demonstrate that single-layer (AO or BO$_2$) deposition following co-deposition is possible. Thus, interface termination can be realised. In Fig.~\ref{fig_01}\textbf{d}, the BSO-related normalized RHEED intensity of the specular spot during approaches (i) vs. (ii) depicted in Fig.~\ref{fig_01}\textbf{a} and~\ref{fig_01}\textbf{b}, respectively, is monitored as a function of BSO growth time. 
 The yellow shaded area marks the final seconds of the BSO growth of an $\approx$ 100nm thick BSO film on a DyScO$_3$ substrate by adsorption-controlled co-deposition (Ba and SnO shutters both open). The constant RHEED intensity likely indicates a stable, mixed BaO and SnO$_2$ termination averaged over the macroscopic footprint of the RHEED beam on the growth front. Starting at $t=10$~s (purple shaded area), either the Ba shutter was closed (whereas the SnO shutter remained open) for SnO$_2$ termination (magenta curve, approach (i)), or the SnO shutter was closed (Ba shutter remained open) for BaO termination (blue curve, approach (ii)). Growth of SnO$_2$ on top (only SnO shutter open, red curve) leads to a decreasing RHEED intensity. Growth of BaO on top (only Ba shutter open, cyan curve), in contrast, leads to an increasing RHEED intensity in accordance with RHEED oscillations of this system \cite{prakash_2015}. The brown shaded area marks the region where both cell shutters were closed to stabilize the surface before continuing with LIO deposition that starts either with one monolayer LaO (approach (i)) or InO$_2$ ((approach (ii)). The continuously constant RHEED intensity after the intended species deposition suggests that these terminations of SnO$_2$ and BaO are stable and remain on the surface. Growth details and discussion on deposition times can be found in the experimental section. 

\subsection{Structural integrity}\label{subsec_struct}

Scanning transmission electron microscopy (STEM) images verify the quality of the BSO/LIO heterostructure.
On a larger length scale (Fig.~\ref{fig_01}\textbf{e}), the DSO substrate, BSO layer, and LIO layers can be clearly distinguished. In addition, misfit dislocations at the DSO/BSO interface are revealed by their strain field. They arise from the significant lattice mismatch between DSO and BSO of $\approx$~4.5$\%$ and result in threading dislocations within the BSO film that penetrate the LIO and likely limit the achievable electron mobility in the heterostructure. A coherent BSO/LIO interface region is shown in Fig.~\ref{fig_01}\textbf{f}. The coherent growth of BSO/LIO heterostructure using approach (i) was also monitored by \textit{in-situ} RHEED [see Supplementary Information (SI) Fig.~\ref{fig_ED_RHEED}]. 

Fig.~\ref{fig_SI_DSO_sub}\textbf{c} and Fig.~\ref{fig_SI_DSO_sub}\textbf{d} of the SI show the surface morphology measured by atomic force microscopy of the corresponding LIO/BSO heterostructure with nominally SnO$_2$/LaO interface [approach (i)] and confirm single crystalline growth of individual BSO as well as LIO layers by XRD scans, respectively. The surface exhibits information of substrate terrace formation (SI Fig.~\ref{fig_SI_DSO_sub}\textbf{a} and \ref{fig_SI_DSO_sub}\textbf{b}), despite a total layer thickness of 130~nm. The heterostructures grown with the BaO/InO$_2$ termination by approach (ii) and without intentional interface termination by approach (iii) (SI Figs.~\ref{fig_SI_afm_xrd_other_IF_term} and ~\ref{fig_SI_tem_no_IF}) show similar structural properties.

\subsection{Interface-termination design determines presence of 2DEG}\label{HX-PES}
\begin{figure*}[t]
\centering
\includegraphics[width=0.75\textwidth]{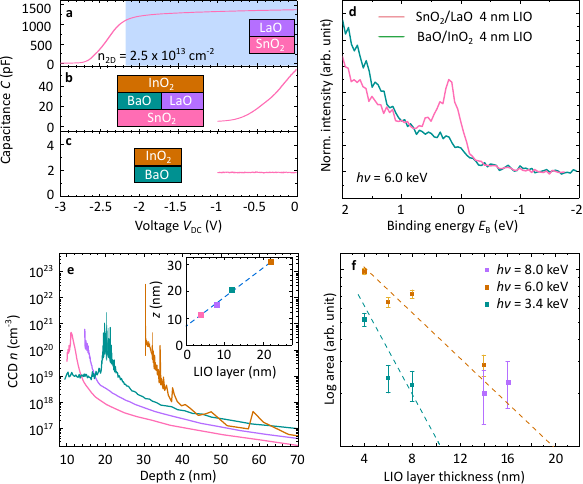}
\caption{ 
Capacitance-Voltage (C-V) measurement of the samples with \textbf{a} SnO$_2$/LaO interface termination, \textbf{b} no interface, and \textbf{c} BaO/InO$_2$ interface termination as depicted in the sketch. From the blue shaded area in \textbf{a}, the charge carrier accumulation at the interface was derived. \textbf{d} Hard X-ray photoelectron spectroscopy (HAXPES) valence band spectra, with the spectra magnified on the Fermi edge to elucidate the charge carriers at the interface for SnO$_2$/LaO (red) and BaO/InO$_2$ (cyan) interface termination for the samples with a 4~nm thick LIO layer. \textbf{e} Charge carrier density (CCD) profiles across the LIO/BSO interface derived from C-V measurements as shown in \textbf{a} using Eq.~\ref{eq_C-V_dept_profile} for different LIO layer thicknesses. Inset: position of the charge accumulation layer as a function of nominally deposited LIO layer thickness.  \textbf{f}  Area dependence of HAXPES charge carrier signal at the Fermi edge shown in \textbf{d} as a function of LIO layer thickness for different photon energies $h\nu$ - all samples contain the SnO$_2$/LaO interface termination. The dashed cyan and brown lines depict the expected signal reduction due to LIO layer thickness and were derived from Sn~3\textit{d}$_{5/2}$ core levels collected at 3.4~keV and 6.0~keV, respectively.}\label{fig_02}
\end{figure*}

Next, we show that the engineered interface terminations discussed above each result in drastically different electrical behaviour of the BSO/LIO samples. 
Capacitance-voltage (C-V) measurements, as described in Ref.~\cite{pfutzenreuter2022} and schematically illustrated in Fig.~\ref{fig_SI_3}\textbf{a} and~\ref{fig_SI_3}\textbf{b}, were performed on BSO/LIO heterostructures grown by both approaches (i) and (ii).

While for the intended SnO$_2$/LaO termination [(approach (i)]  the measurements, shown in Fig.~\ref{fig_02}\textbf{a}, indicate charge carrier accumulation in the range of low 10$^{13}$ cm$^{-2}$ (blue shaded area), any mobile charge carriers are completely absent for the intended BaO/InO$_2$ interface termination realized by approach (ii), as shown in Fig.~\ref{fig_02}\textbf{c}. For BSO/LIO heterostructures without intentionally defined interface termination [(approach (iii)], the C-V measurements indicate sample-to-sample variation fitting between the two above-discussed scenarios (Fig.~\ref{fig_02}\textbf{b} and SI Fig.~\ref{fig_SI_3}\textbf{c}).

These results are independently corroborated by hard X-ray photoelectron spectroscopy (HAXPES) measurements. The signal at the Fermi energy, i.e., at binding energy $E_B=0$, in Fig.~\ref{fig_02}\textbf{d} indicates free charge carriers in the BSO/LIO system. The recorded free carrier signal of a sample grown with shutter-controlled SnO$_2$/LaO interface termination (magenta solid line, approach i) is always larger compared to the signal from BaO/InO$_2$ interface termination (cyan solid line, approach ii). Details on HAXPES measurements can be found in the experimental section as well as in the SI and related Figs.~\ref{fig_SI_3} and ~\ref{fig_SI_haxpes_survey}.

C-V and HAXPES experiments of SnO$_2$/LaO-terminated BSO/LIO heterostructures grown by approach (i) with varying LIO layer thickness further confirm that the charge carrier accumulation is located at the interface.
The peak position of the volume carrier density as a function of probing depth, shown in Fig.~\ref{fig_02}\textbf{e} and obtained from the C-V measurements using Eq.~\ref{eq_C-V_dept_profile}, reflects the distance of the charge carrier accumulation to the LIO surface. As shown in the inset of Fig.~\ref{fig_02}\textbf{e}, it follows the nominal LIO layer thickness, suggesting accumulation at the LIO/BSO interface. A $\delta_{\text{off}}$~=~7~nm offset was applied to the data, which we attribute to a dielectric contamination layer accumulated at the Hg surface. \newline 
In HAXPES measurements, the same results were achieved using different X-ray energies in combination with different LIO layer thicknesses from 4~nm to 20~nm, shown in Fig.~\ref{fig_02}\textbf{f}. The HAXPES spectra from which the area values at E$_f$ were extracted are shown in SI Fig.~\ref{fig_ED_haxpes}.  As the photon energy is increased from 3.4~keV to 8.0~keV, the maximum inelastic mean free path (IMFP) of the photoelectrons increases from 4.95~nm to 10.1~nm \cite{tanuma2005}. The probing depth is generally considered to be three times the IMFP (equivalent to detecting 95\% of the total photoelectron signal), allowing for the detection of charge carrier accumulation at the BSO/LIO interface through thicker LIO layers. 

Only when the photon energy was high enough to excite photoelectrons from the buried interface, is it possible to detect the charge accumulation feature in the HAXPES data. 

Since the signal of charge carriers at the $E_F$ decreases with increasing LIO layer thickness in the same manner as the signal of Sn 3$d$ core-level (cyan and brown dashed lines at 3.4~keV and 6.0~keV, respectively), we attribute the presence of the charge carriers to the interface of the BSO/LIO heterostructure. We rule out the presence of a broader doped region inside the BSO layer by comparison to Fig.~\ref{fig_02}\textbf{d}, which compares two identical heterostructures which only differ in their interface termination.


Interestingly, the C-V measurements also demonstrate that full depletion of the charge accumulation is possible - a mandatory ingredient for further processing towards transistor applications. Moreover, applied voltages of 3~V to 10~V for LIO layer thicknesses in the range of 4~nm to 24~nm correspond to high breakdown field strengths of the LIO layers in the range of 2.7~-~3.2~MV/cm (considering the $\delta_{\text{off}}$~=~7~nm offset).

\subsection{2DEG transport properties}\label{subsec_el_char}

\begin{figure*}[t]
\centering
\includegraphics[width=0.75\textwidth]{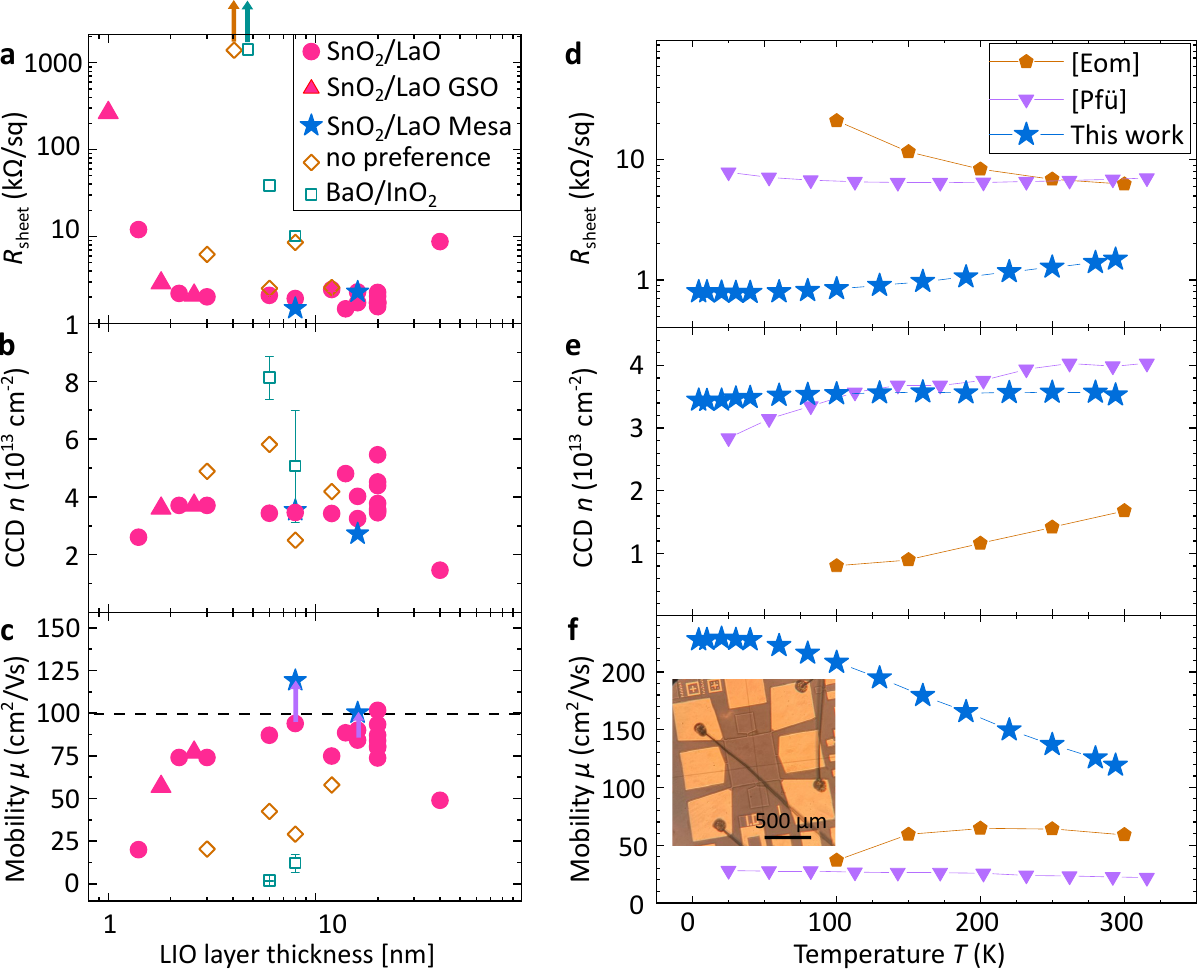}
\caption{
Electrical characterization of \textbf{a} sheet resistance, \textbf{b} CCD, and \textbf{c} mobility as a function of LIO layer thickness and interface termination at room temperature. \textbf{d--f} temperature-dependent measurement and comparison to other BSO based 2DEGs labeled as follows: [Eom]\cite{eom2022}, [Pfü]\cite{pfutzenreuter2022} }\label{fig_03}
\end{figure*}

The electron transport properties in the as-grown heterostructures with different LIO layer thicknesses and designed interface termination were determined by van-der-Pauw-Hall (vdP) measurements. Details on contact deposition and lithographically defined structures (Fig.~\ref{fig_03}\textbf{f}) can be found in the experimental section. 
In Fig.~\ref{fig_03}~\textbf{a--c}, the RT sheet resistance, charge carrier density, and mobility, respectively, are shown for different LIO layer thicknesses, different substrates, and differently designed interface terminations.

We find that the heterostructures with designed SnO$_2$/LaO interface (magenta circles) generally exhibit the lowest sheet resistance values ($\approx$~2~k$\Omega/\square$) and charge carrier densities of $\approx$~3.5~-~5~$\times$~10$^{13}$~cm$^{-2}$, resulting in mobility values in the range of 75~-~100 cm$^2$/Vs for LIO thicknesses from 2~-~20~nm. In addition, similar values were achieved using GdScO$_3$ (GSO) substrates (magenta triangles), demonstrating that our approach is not limited to the choice of substrate.
The increase of sheet resistance and drop of CCD and mobility for heterostructures with decreasing LIO layer thickness below 2~nm is in accordance with theoretical predictions \cite{aggoune2021} corresponding to ionic displacement and octahedral tilt in both materials that screen the electric field induced by the polar discontinuity. Such critical thickness for 2DEG formation has also been observed in PLD-grown samples \cite{kim2018}.
Using a lithographically-defined, mesa-isolated vdP structure, record mobilities in perovskite oxide 2DEGs up to 119~cm$^2$/Vs at RT were measured (blue asterisks in Fig.\ref{fig_03}\textbf{a--c}), which is by far the highest value for as-grown samples reported in the literature to the best of our knowledge.

\begin{figure*}[t]
\centering
\includegraphics[width=0.5\textwidth]{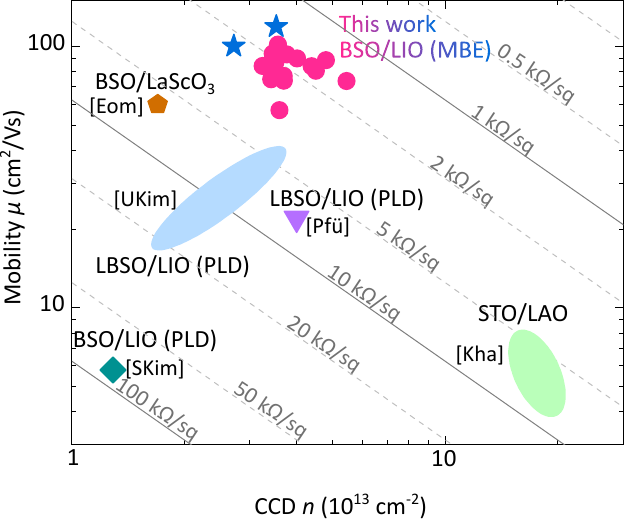}
\caption{ Comparison of RT 2DEG sheet resistance, charge carrier density, and mobility to other oxide systems. The reference data is indicated by square brackets with the following relation: [Kal]\cite{kalarickal2020}, [Eom]\cite{eom2022}, [UKim]\cite{kim2016}, [Pfü]\cite{pfutzenreuter2022}, [SKim]\cite{kim2023a}, [Kha]\cite{khan2017}.}\label{fig_04}
\end{figure*}

In contrast, significantly higher sheet resistances were measured for samples grown with the designed BaO/InO$_2$ interface (empty cyan squares). Hall measurements indicate an increased charge carrier density and, thus, low mobilities. The high apparent Hall charge carrier concentration contradicts the results of C-V measurements. It is considered to be an artefact that results from non-ideal transport along perculative paths causing a drop in measured Hall-voltage. This is further elaborated in the SI and related Fig.~\ref{fig_SI_cheese_model}.

Electron transport properties of BSO/LIO heterostructures without designed interface termination (empty brown diamonds) fluctuate between those of heterostructures with designed SnO$_2$/LaO interface termination and BaO/InO$_2$ interface termination in line with the C-V measurements shown in Fig.~\ref{fig_SI_3}~\textbf{c} (see SI), and are thus consistent with an uncontrolled, fluctuating, possibly laterally inhomogeneous/mixed interface termination for these samples.\newline 

The temperature dependence of transport properties of a SnO$_2$/LaO interface terminated 120~nm/20~nm BSO/LIO sample shown in Fig.\ref{fig_03}\textbf{d--f} qualitatively agrees with the behaviour of other known 2DEGs: A decreasing sheet resistance with decreasing temperature from 300~K to 4~K can be observed in Fig.\ref{fig_03}\textbf{d}. The comparably constant electron concentration in this temperature range, visible in Fig.\ref{fig_03}\textbf{e}, is a further indication of charge accumulation (a degenerate system that cannot freeze out). 
The related electron mobility in Fig.~\ref{fig_03}\textbf{f} increases with decreasing temperature, in agreement with the reduction of phonon scattering, and saturates at 228~cm$^2$/Vs below 50~K -- likely due to the presence of threading dislocations originating from the DSO/BSO interface.
The temperature-dependent transport properties (cf. Fig.~\ref{fig_03}\textbf{d--f}) of previously published work show significantly lower electron mobilities (Ref.~\cite{pfutzenreuter2022}) or indication of a freeze-out of the carrier in Ref.~\cite{eom2022}, which is atypical for a real 2DEG.

Hence, the MBE-grown LIO/BSO samples of the present work with a shutter-controlled SnO$_2$/LaO interface termination stand out by a factor of two or more compared to published RT transport properties of other BSO-related 2DEGs (Fig.~\ref{fig_04}) in terms of high CCD, high mobility and low sheet resistance (both crucial for HEMT applications). STO-based systems, that can have higher charge carrier density are clearly outperformed in terms of application-relevant RT electron mobility.
\section{Summary and conclusion}\label{sec:summary}
In summary, we have experimentally identified the interface termination as a key ingredient for high-mobility 2DEGs in BSO/LIO heterostructures. We have furthermore established a path to realising a specific interface termination while ensuring high layer quality for MBE-grown BSO/LIO samples by combining an adsorption-controlled co-deposition with a layer-by-layer growth of the interfacial monolayers.
Both C-V and HAXPES measurements show a drastically enhanced charge accumulation at the engineered BSO/LIO interface with SnO$_2$/LaO interface termination, in agreement with theoretical predictions \cite{aggoune2021}.
In such samples, electrical transport measurements revealed room temperature mobility values up to 119~cm$^2$/Vs, sheet resistance values down to 1.8 k$\Omega/\square$ and CCD of $n \approx$~3.6~$\times$~10$^{13}$~cm$^{-2}$.

Further mobility improvement can be expected for samples with reduced dislocation density and improved interface roughness. Notwithstanding, our results already surpass existing BSO-based 2DEGs by a factor of at least two in terms of increased mobility and decreased sheet resistance. We further note that the high breakdown field strengths in the range of 2.7~-~3.4~MV/cm in the LIO layer make it a suitable dielectric that allows full depletion of the 2DEG. All this, in combination with high charge carrier densities, establishes BSO/LIO 2DEGs as promising building blocks for perovskite-based HEMTs with unprecedented performance. Additionally, their potential for monolithic integration with further perovskite-based functional layers opens up for realization of novel oxide electronic devices such as high-electron mobility ferroelectric FETs \cite{fang2023,park2020,kim2023}. Lastly, this approach of combining co-deposition with layer-by-layer growth to realise a specific interface termination is not a unique feature of the BSO/LIO heterostructure but can be applied to any other perovskite system.

\section{Experimental methods} \label{sec:experiment}

\textit{MBE growth:}\newline
For the growth of the BSO/LIO heterostructure, DyScO$_3$ (DSO) substrates were used. Details on substrate preparation and LIO growth can be found elsewhere \cite{hoffmann2022}.
The BSO was grown using a mixture of SnO$_2$ and Sn as source material resulting in a SnO suboxide flux at comparably low cell temperatures in the range of 740$\thinspace^{\circ}$C~-~850$\thinspace^{\circ}$C. Details about the suboxide sources can be found elsewhere \cite{hoffmann2020}.
For the present study, Ba, SnO, In, and La cell fluxes were chosen in such a way, that substrate temperature, oxygen flux and plasma power were kept constant at 835$\thinspace^{\circ}$C (measured by a pyrometer), 0.065~sccm, and 200~W, respectively. To keep the plasma source running, 0.2 sccm Ar supporting gas was added to the gas flow. More details can be found in the SI.
As a consequence of the adjustment of the B-cation fluxes and growth parameters, the growth of the BSO/LIO heterostructure and interface termination reduces to a simple cell shutter sequence.\newline

\textit{Interface termination}\newline
Assuming that the BSO growth front during Co-deposition exhibits a SnO$_2$/BaO termination ratio in the range of 1:2~-~2:1. To achieve full coverage with the same termination up to 0.67~ML of the intended species needs to be supplied. For a BSO growth rate of 1.8~nm/min, the time for a BaO or SnO$_2$ layer to form would be $\approx$~14~s. Consequently, shutter times in the range of 9-10~seconds seem to be suitable for the realization of majority terminations. For The LIO growth, LaO or InO$_2$ shutter was opened for $\approx$ 15~s (a growth rate of 1.2~nm/min corresponds to LaO or InO$_2$ monolayer deposition within 20~s. Thus, 3/4 of the interface termination is defined before starting co-deposition growth that completes interface termination and realizes LIO growth.
\newline
 
\textit{TEM:}\newline
For TEM analysis of the samples, an aberration-corrected FEI Titan 80-300 operating at 300~kV was used. For recording scanning TEM (STEM) images, a high-angle annular dark-field (HAADF) detector and a camera length of 195~mm was used. TEM cross-sectional samples were prepared and analyzed along the [110] projection of the DSO substrate.
Analyzed heterostructures are projected along the [110] direction of DSO and [100] direction of BSO.\newline

\textit{Processing transport structures}\newline

For electrical characterization, Ti/Au contacts with a thickness of 20/100~nm were deposited at the corners (as well as at the edges) of the 5$\times$5~mm$^2$ samples by sputtering using a shadow mask. The diameter of the contacts was 0.5~mm at the edges and 1~mm at the corners. \newline 
Further, structures with a more precise geometry were realized by photolithography using a shadow mask. For mesa definition, wet chemical etching was done (2~min sample etching in HF:HNO$_3$:H$_2$O$:2$ / 10:4:100~mL at an etching rate of 70~nm/min).
For sample contacts, Ti/Au 10/90~nm were deposited using an electron beam evaporator. For the lift-off of metal on resist acetone was used.\newline
\textit{
Electrical transport:}\newline
The samples were electrically investigated using the vdP method \cite{vanderpauw1958,vanderpauw1958a}. The following formulas were used to extract sheet resistance $R_{\text{sheet}}$ and charge carrier density $n=\frac{1}{eR_H}$, from which mobility $\mu=\frac{R_H}{\rho}$ was derived:
\begin{align}
R_{\text{sheet}} = \frac{\pi}{\ln 2}\frac{R_{\text{(AB,DC)}}+R_{\text{(BC,AD)}}}{2}\cdot f_{\text{vdP}}. \label{eq_rho_vdp}
\end{align}
The 4-terminal resistance is defined by $R_{(AB,DC)}=I_{AB}/U_{DC}$ (Fig.~\ref{fig_03}a), and $f_{\text{vdp}}$ is a form factor that takes the sample geometry into account, and that can be derived from the expression \cite{vanderpauw1958,vanderpauw1958a}:
\begin{align}
\cosh {\left\lbrace  \frac{R_{\text{(AB,DC)}}/R_{\text{(BC,AD)}}-1}{R_{\text{(AB,DC)}}/R_{\text{(BC,AD)}}+1}\cdot \frac{\ln 2}{f_{\text{vdP}}}\right\rbrace }=\frac{1}{2}\cdot \exp \frac{\ln 2}{f_{\text{vdP}}}.\label{eq_vdp-formfactor}
\end{align}
Note that Eq.~\ref{eq_vdp-formfactor} can only be solved implicitly.\\
Further, the Hall resistance in vdP configuration is given by \cite{vanderpauw1958,vanderpauw1958a}:
\begin{align}
R_H=\frac{1}{B_z}\Delta R_{(BD,AC)},\label{eq_Rhall_vdp}
\end{align}
where $\Delta R_{(BD,AC)}$ is the resistance change of $R_{(BD,AC)}$ due to the magnetic field $B_z$ as shown in Fig.~\ref{fig_04}(b). 

\textit{Capacitance-voltage measurements: }\newline
C–V measurements were performed using a Keithley 4200 measurement device connected to an Hg probe system with a 0.3~mm diameter disc-shaped gate contact. A 10~kHz, 50~mV rms AC voltage was used and the capacitance was extracted using the series circuit model.
Figs.~\ref{fig_SI_3}(a) and ~\ref{fig_SI_3}(b) show a top and side view of a LIO/BSO heterostructure with two mercury probes being in contact with the sample surface, respectively. In Fig.~\ref{fig_SI_3}(b) the equivalent circuit model is drawn into the depicted LIO/BSO heterostructure. Since the capacitance $C_2 \gg C_1$ due to the different area size (the area of $C_1$ has a size of $A_1=7.3\times10^{-2}$~mm$^2$, the area of $C_2$ has a size of $A_2\approx 19 \times A_1$), the total capacitance is given by:
\begin{align}
\frac{1}{C}=\frac{1}{C_1}+\frac{1}{C_2} \approx \frac{1}{C_1}.\label{eq_C-V_total_capacitance}
\end{align}
The samples were analysed using a serial model at an AC frequency in the range of 10-50~kHz and using $\epsilon _r$ of 21 for the calculation of the capacitance.\\
In addition, the 2D-charge carrier density can be calculated according to:
\begin{align}
n_{2D}=\frac{C \Delta V_{\text{dc}}}{A_1 e},\label{eq_C-V_n2d}
\end{align}
where $\Delta V_{\text{dc}}$ is the $dc$ voltage difference for which $C$ is constant (blue shaded area in Fig.~\ref{fig_03}c), and $A_1$ is the size of the small Hg contact.\\
The CCD as a function of the depth can be derived from the slope of $1/C^2$ as a function of $V_{\text{dc}}$: 
\begin{align}
n(z)\mid _{z=W}=-\frac{2}{\epsilon_r \epsilon_0}\left[\frac{d\left(\frac{1}{C^2}\right)}{dV_{\text{dc}}}\right]^{-1}.\label{eq_C-V_dept_profile}
\end{align}
Note that the relation between $C$ and $W$ is given by:
\begin{align}
C=\frac{\epsilon_r \epsilon_0 A}{W}.\label{eq_C-W_relation}
\end{align}

\begin{figure*}[t]
    \centering
    \includegraphics[width=0.99\linewidth]{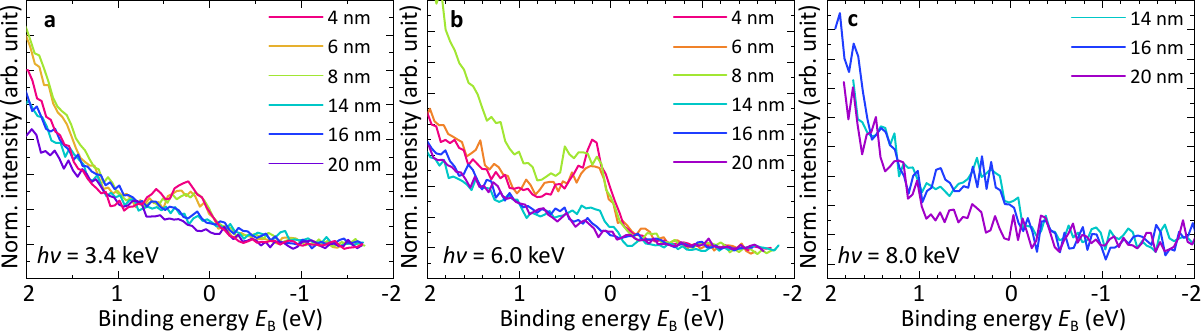}
    \caption{HAXPES measurements of charge carriers at the Fermi edge for different photon energies $h\nu$ as well as LIO layer thicknesses - all samples contain the SnO$_2$/LaO interface termination.
    HAXPES scans around the Fermi edge of BSO/LIO heterostructures at photon energies of \textbf{a} 3.4~keV, \textbf{b} 6.0~keV and \textbf{c} 8.0~keV. The legend depicts the LIO layer thickness of the samples. The signal of free charge carriers is measured as an increase in the intensity of the states below the Fermi edge. With an increase in the photon energy, the LIO layer thickness through which the charge carriers can be detected increases from 8~nm at 3.4~keV to 16~nm at 8.0~keV, thus demonstrating that the charge carriers are not a surface effect but located at the BSO/LIO interface.}
    \label{fig_ED_haxpes}
\end{figure*}

\textit{Hard X-ray photoelectron spectroscopy (HAXPES):} \newline
HAXPES data were collected at beamline P22 at PETRAIII, German Electron Synchrotron DESY in Hamburg, Germany~\cite{Schlueter2019}. Three photon energies were used, including 3.4, 6.0 and 8.0~keV. A Si (311) double-crystal monochromator was used to achieve the two higher photon energies. A double channel-cut post-monochromator employing a Si(111) and a Si(220) channel-cut crystal pair was used to select 3.4~keV. The beamline employs a Phoibos 225HV analyzer (SPECS, Berlin, Germany), which was set up in the small area lens mode with a slit size of 3~mm for the collection of spectra. Spectra were collected using a pass energy of 30~eV. The total energy resolution for the three photon energies in this setup was determined based on the 16/84\% Fermi edge ($E_F$) width of a polycrystalline gold foil \cite{Kalha2023}. The resulting total resolutions were 215~meV, 249~meV, and 260~meV for 3.4~keV, 6.0~keV and 8.0~keV, respectively. All experiments were conducted in grazing incidence geometry ($\leq$5$^\circ$). Survey scans and core-level (CL) spectra are shown in SI Fig.~\ref{fig_ED_haxpes}. From all recorded CL spectra, a Shirley background was subtracted. The spectra were normalised to the sum of the Sn 3$d_{5/2}$ and In 3$d_{5/2}$ core-level area and either aligned to the internal 50$\%$ value at the Fermi edge or to $E_F$ (Au) in case of absence of free carriers at $E_F$. For the investigation of the 2DEG area, an additional linear background was subtracted.

\newpage
\section{Supplementary Information}

Supplementary Information is available at...

\section{acknowledgments}
We thank Hanjong Paik and Darrell Schlom for helpful discussion on BSO growth, Roger de Souza, who helped us identify contaminations in the early stage of BSO growth, Walid Anders, Sander Rauwerdink, and Nicole Volkmer for processing the samples, as well as Roman Engel-Herbert for fruitful discussions. This work was partially performed in the framework of GraFOx, a Leibniz ScienceCampus partially funded by the Leibniz association. G.H. and M.Z. gratefully acknowledge financial support from the Leibniz-Gemeinschaft under Grant No. K74/2017. We acknowledge DESY (Hamburg, Germany), a member of the Helmholtz Association HGF, for the provision of experimental facilities. Parts of this research were carried out at PETRA III using beamline P22  for proposal I-20221263. Funding for the HAXPES instrument by the Federal Ministry of Education and Research (BMBF) under framework program ErUM is gratefully acknowledged.

\section{Conflict of interest}
The authors declare no conflict of interest
\section{Data Availability Statement}
Data are available from the corresponding author upon reasonable request.

\bibliographystyle{wb_stat}
\bibliography{WileySTAT}
\newpage\hbox{}\thispagestyle{empty}
\section{Supplementary Information}
\begin{figure*}[t]
    \centering
    \includegraphics[width=0.99\linewidth]{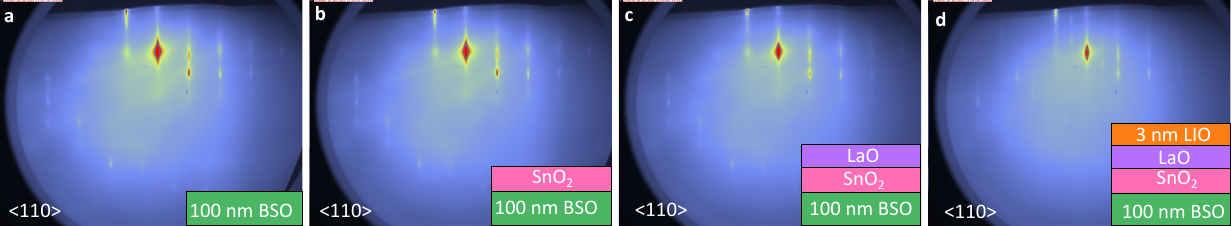}
    \caption{\textbf{a--d} RHEED images indicating continuity of the crystal structure during shutter-controlled interface formation:
\textbf{a} RHEED image along the BSO <011> azimuth of a 100~nm thick BSO film grown on a DyScO$_3$ substrate using Ba and SnO co-deposition. \textbf{b} RHEED image of nominally SnO$_2$ terminated BSO surface by closing the Ba shutter 10 s before the SnO shutter. \textbf{c} Realizing of SnO$_2$/LaO interface termination by opening only the La shutter 15~s before the In shutter. \textbf{d} LIO growth using co-deposition of La and In in an adsorption-controlled growth regime. }
    \label{fig_ED_RHEED}
\end{figure*}
\newpage\hbox{}\thispagestyle{empty}\newpage
\subsection*{Growth of BaSnO$_3$ and LaInO$_3$}
BSO and LIO are grown in an adsorption-controlled growth \cite{paik2017,hoffmann2022}, i.e. growth in excess of the more volatile B-cation. However, for both LIO and BSO, the growth window is rather narrow ($\pm$~2$\thinspace^{\circ}$C in the cation fluxes).
The Ba, La, and In cell temperatures were in the range of 580$\thinspace^{\circ}$C~-~630$\thinspace^{\circ}$C, 1490$\thinspace^{\circ}$C~-~1550$\thinspace^{\circ}$C, and 700$\thinspace^{\circ}$C~-~705$\thinspace^{\circ}$C, respectively. The resulting growth rates of BSO and LIO were in the range of 1.8-2.1~nm/min, and 1.1-1.4~nm/min. Note that SnO and In were given in a slight excess (ratio of 1/1.5~-~1/3 to Ba and La, respectively) to ensure the adsorption-controlled growth. 

\subsection*{Substrate preparation and sample analysis}
\begin{figure*} [t]
\centering
\includegraphics[width=0.75\textwidth]{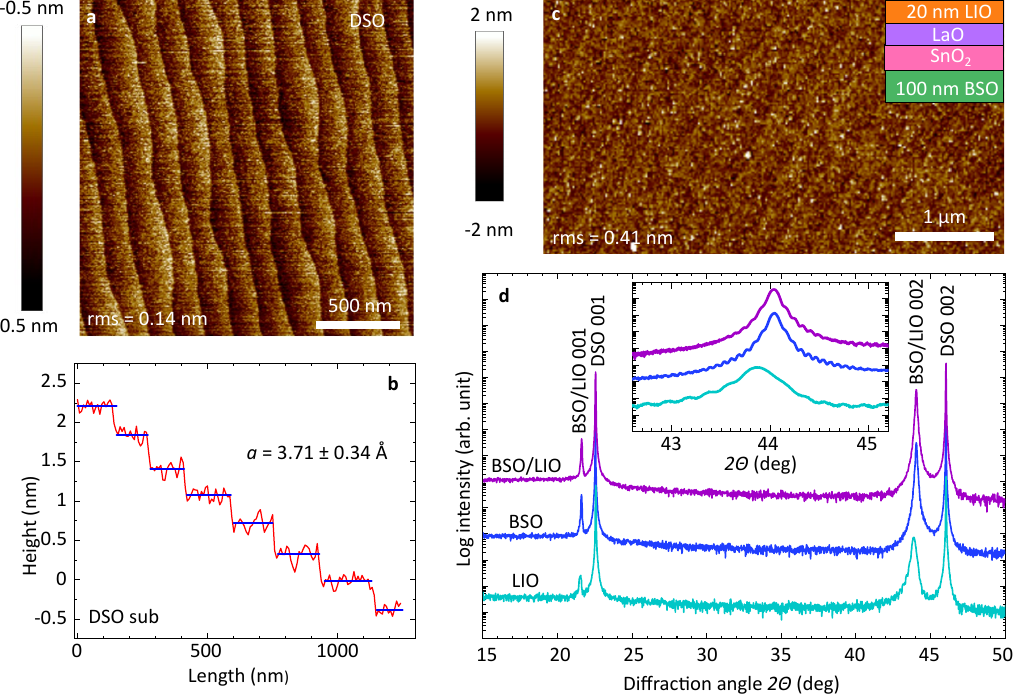}
\caption{\textbf{a} DSO substrate surface after preparation procedure. \textbf{b} shows the corresponding height profile derived from line scans. The average step edge height indicates single surface termination.\textbf{c} Atomic force microscopy image of the grown heterostructure and interface design as depicted in the sketch in the upper right corner. \textbf{d} X-ray diffraction symmetric $2\Theta-\omega$ scan for a grown LIO/BSO heterostructure (16/110~nm) with interface design as well as for single BSO (110~nm) and LIO (60~nm) films, all grown on DSO substrates. Thickness fringes of the 002 reflex shown in the inset reveal the presence of sharp interfaces.}\label{fig_SI_DSO_sub}
\end{figure*}

The DSO samples were annealed in a tube furnace with a quartz tube at 1050$\thinspace^{\circ}$C for 6~h. Additionally, an oxygen gas flow of $\approx$ 250~sccm was applied for the reconstruction of the DSO surface. The temperature ramp rates were set to 20$\thinspace^{\circ}$C/min for heating up and cooling down. Fig.~\ref{fig_SI_DSO_sub}\textbf{a} shows the DSO surfaces after the substrate preparation. It is assumed that the surfaces are singly terminated as indicated by the profiles of DSO shown in Fig.~\ref{fig_SI_DSO_sub}\textbf{b}. According to the chosen annealing parameters for DSO, ScO$_2$ (B-site) termination is assumed \cite{dirsyte2010}.\\

\begin{figure*}[t]
    \centering
    \includegraphics[width=0.6\linewidth]{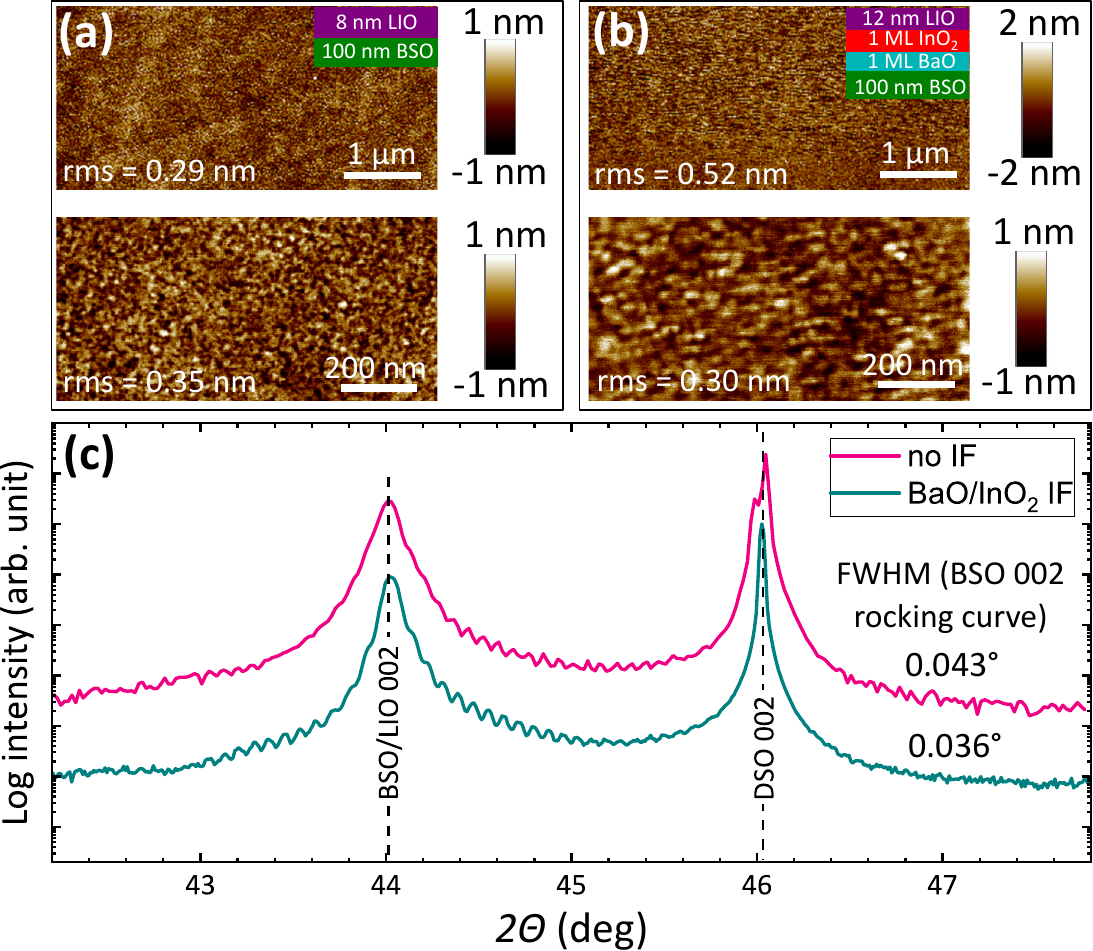}
    \caption{AFM images of a sample without specific interface termination \textbf{a} and with BaO/InO$_2$ interface termination \textbf{b}. \textbf{c} XRD 2$\Theta - \omega$ scan showing the BSO/LIO 002 reflection and the DSO substrate 002 reflection. Thickness fringes and FWHM values of the BSO 002 rocking curve indicate high crystalline quality.}
    \label{fig_SI_afm_xrd_other_IF_term}
\end{figure*}

\begin{figure*}[t]
\centering
\includegraphics[width=0.75\textwidth]{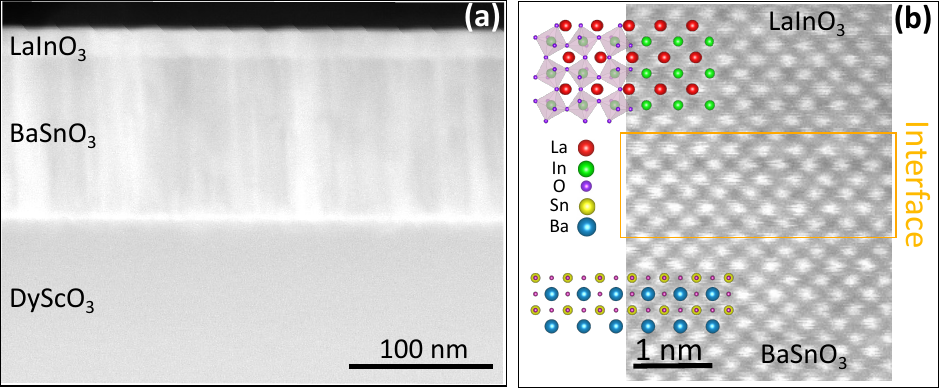}
\caption{STEM images of a LIO/BSO/DSO heterostructure \textbf{a} and coherent LIO/BSO interface \textbf{b} without any preferential interface termination. Atomistic models of LIO and BSO were superimposed on the image. The orange frame indicates the transition region from BSO to LIO.}\label{fig_SI_tem_no_IF}
\end{figure*}

Figs.~\ref{fig_SI_afm_xrd_other_IF_term}\textbf{a} and~\ref{fig_SI_afm_xrd_other_IF_term}\textbf{b} show AFM images of BSO/LIO heterostructures without any preferential interface termination and BaO/InO$_2$ interface termination, respectively. The root mean square (rms) values are in the same range as the ones for SnO$_2$/LaO terminated heterostructures. Further, Fig.~\ref{fig_SI_afm_xrd_other_IF_term}\textbf{c} shows a symmetric XRD 2$\Theta - \omega$ scan of the corresponding samples in Fig.~\ref{fig_SI_afm_xrd_other_IF_term}\textbf{a} and~\ref{fig_SI_afm_xrd_other_IF_term}\textbf{b}. Also here, thickness fringes and FWHM values of BSO 002 rocking curves are comparable to the ones for SnO$_2$/LaO interface termination. In addition, Fig.~\ref{fig_SI_tem_no_IF} shows a STEM image of a BSO/LIO heterostructure that has no specific interface termination. As for the STEM analysis of the sample with a SnO$_2$/LaO interface termination in the main paper, the crystallinity of the BSO and LIO layers (Fig.~\ref{fig_SI_tem_no_IF}\textbf{a}) as well as the coherent growth of LIO on BSO is confirmed (Fig.~\ref{fig_SI_tem_no_IF}\textbf{b}). Since the morphology of the samples grown according to approaches (i), (ii), and (iii) depicted in Fig.~\ref{fig_01}\textbf{a} is almost identical, we exclude the hypothesis, that variations of electrical transport properties are caused by morphological changes, but must be a direct consequence of the interface termination.

\newpage

\newpage
\subsection*{Discussion of Hall data}\label{SI_Hall_data}
\begin{figure*}[t]
\centering
\includegraphics[width=0.8\textwidth]{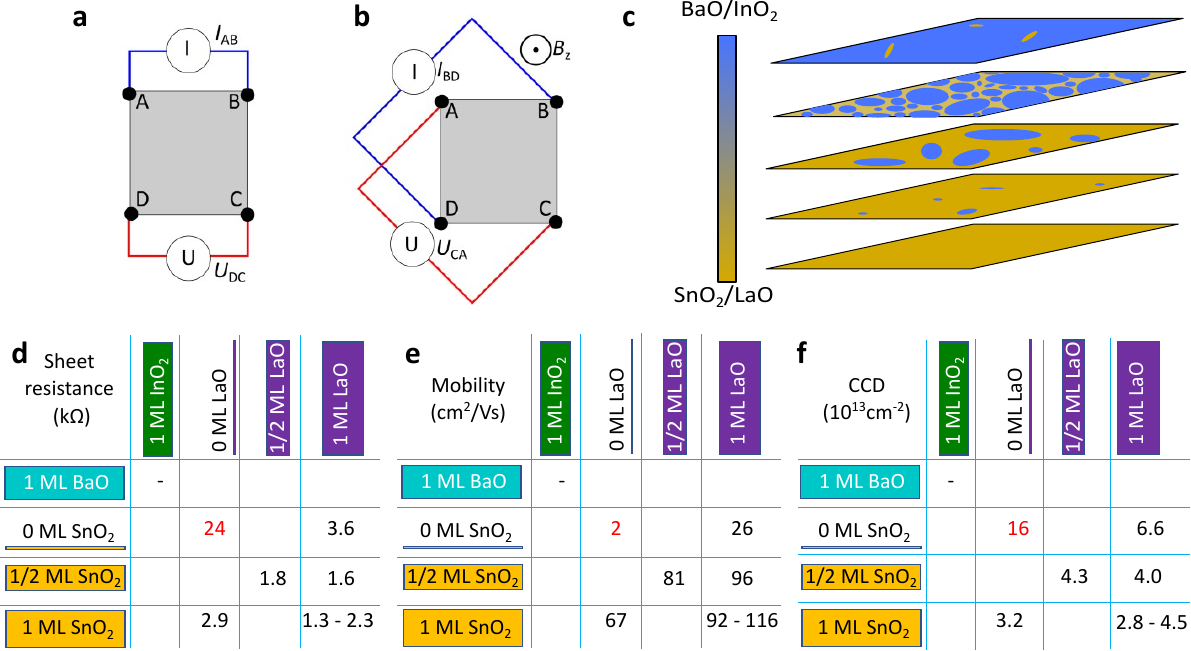}
\caption{Illustration of transition from SnO$_2$/LaO to BaO/InO$_2$ interface termination. Since for areas with BaO/InO$_2$ termination, no electron transport is predicted, longer pathways and additional scattering are present in the case of mixed interface terminations. }\label{fig_SI_cheese_model}
\end{figure*}

Due to finite interfacial roughness and growth conditions, the interface termination can vary between a total SnO$_2$/LaO termination and a BaO/InO$_2$ termination (Fig.~\ref{fig_SI_cheese_model}). While layer-by-layer interface growth tends to be close to single terminated interfaces, the termination contribution is completely different and unpredictable in the case of BSO and LIO co-deposition. 
Assuming still both interface terminations with a minority of the 2DEG hosting SnO$_2$/LaO interface termination, the prerequisites of vdP measurements are not given any more (i.e., homogeneous film without holes). Instead, electrons are now travelling only in the SnO$_2$/LaO interface areas along percolating (and by this longer) pathways to make it to the hall electrodes. Due to the longer pathway, charge carriers are scattered more often and no longer contribute to the Hall signal (a potential difference that only arises because of the presence of charge carrier accumulation at one of the electrodes due to the Lorentz force). While along the applied electric field, this leads to an increased sheet resistance in accordance with the observations, the Hall voltage is reduced resulting in a diminished Hall resistance and artificially increased CCD.
\newpage
\subsection*{Discussion of 2DEG identification}

\begin{figure*}[t]
    \centering
    \includegraphics[width=0.9\linewidth]{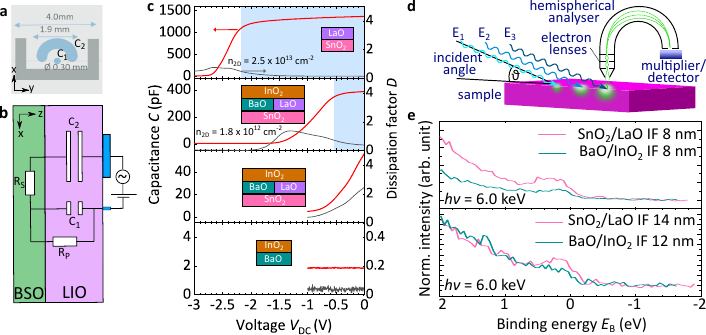}
    \caption{\textbf{a} and \textbf{b} Schematic of contact geometry for C–V measurements from the top and the side, respectively, including equivalent current circuit. Blue marks the Hg contacts, light and dark grey mark the size of the samples and the vacuum pad, respectively. \textbf{c} C-V measurement (solid red line) with respect to the left axis of the samples with different or no specific interface design as depicted in the sketch. From the area within the blue-shaded region, the charge carrier accumulation at the interface can be derived. The dissipation factor of the measurements is shown as a solid, black line with respect to the right axis. \textbf{d} Illustration of X-ray photoelectron spectroscopy (PES) measurements with a variable photon energy source. \textbf{e} Hard X-ray photoelectron spectroscopy (HAXPES) measurements of charge carriers at the interface for SnO$_2$/LaO (red) and BaO/InO$_2$ (green) interface termination for samples with different LIO layer thicknesses [8~nm, and 12(14)~nm].}
    \label{fig_SI_3}
\end{figure*}

Fig.~\ref{fig_SI_3}\textbf{a} and~\ref{fig_SI_3}\textbf{b} give insight into C-V measurement setup. Fig.~\ref{fig_SI_3}
 shows an additional data set for a sample without preferred interface termination (third panel) as well as the Dissipation factor (black solid line) with respect to the right axis. 
 Fig.~\ref{fig_SI_3}\textbf{d} illustrates the HAXPES measurement setup with three different photon energies.
 Fig.~\ref{fig_SI_3}\textbf{e} shows the 2DEG signal at the Fermi edge for samples with SnO$_2$/LaO and BaO/InO$_2$ interfaces for different LIO layer thicknesses written in the legend. Comparison of the data attributes an increased charge carrier accumulation for samples with SnO$_2$/LaO interface.
\newpage
\subsection*{Discussion of HAXPES survey scan and core-level data}

\begin{figure*}[t]
    \centering
    \includegraphics[width=1\linewidth]{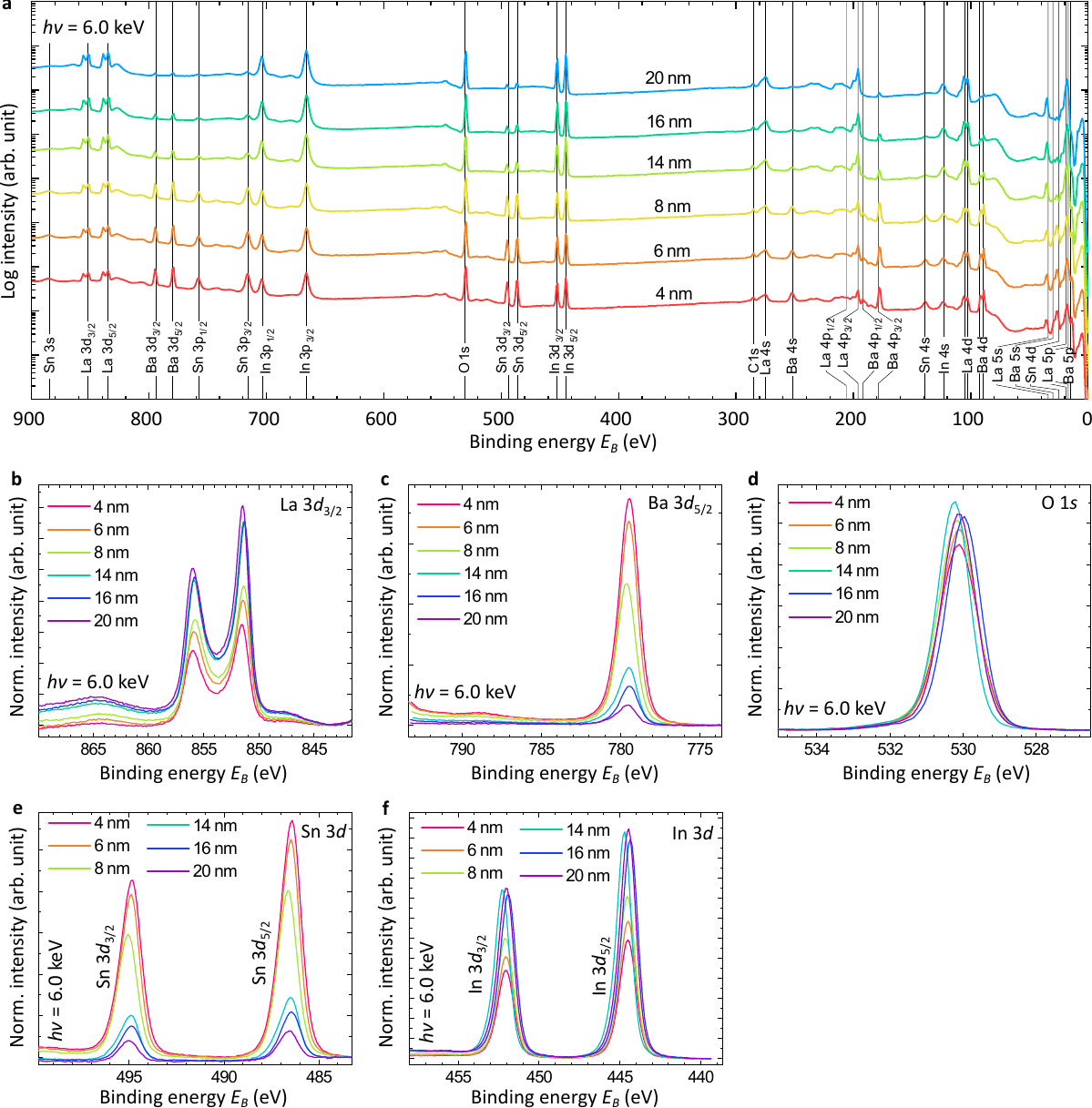}
    \caption{\textbf{a} Representative HAXPES survey spectra of BSO/LIO heterostructures with varying LIO layer thickness from 4~nm to 20~nm written on top of the recorded spectra. The spectra were recorded at 6.0~keV and separated by an offset for better visualisation. Black vertical lines identify the core-level positions. \textbf{b}--\textbf{f} core-level spectra collected at 6.0~keV across the sample range. The spectra displayed are \textbf{b}: La 3\textit{d}\textsubscript{1/2}, \textbf{c}: Ba~3\textit{d}\textsubscript{5/2}, \textbf{d}: O~1\textit{s}, \textbf{e}: Sn~3\textit{d}, and \textbf{f}: In~3\textit{d}.}
    \label{fig_SI_haxpes_survey}
\end{figure*}

Fig.~\ref{fig_SI_haxpes_survey}\textbf{a} shows representative survey scans of samples with SnO$_2$/LaO interface termination at 6.0~keV. The LIO layer thickness is written on top of the individual survey scan. All measured core levels can be attributed to BSO and LIO except a minor peak at 283 eV which corresponds to C 1\textit{s} demonstrating the quality of the samples. The core-level spectra shown in Fig.~\ref{fig_SI_haxpes_survey}\textbf{b}--\textbf{f} reflect three aspects: (I) as LIO layer thickness increases La 3$d_{3/2}$ and In 3$d$ CL intensities increase, (II) with increasing LIO layer thickness, Ba 3$d_{5/2}$ and Sn 3$d$ signal decreases, and (III) O 1$s$ signal remains approximately constant.
\end{document}